# Analysis of effects to scientific impact indicators based on the coevolution of coauthorship and citation networks

Haobai Xue[1]

(xuehb@utszlib.edu.cn; 1. Shenzhen Science & Technology Library/University Town Library of Shenzhen, 2239 Lishui Road, Nanshan District, Shenzhen 518055, China)

**Abstract:** While computer modeling and simulation are crucial for understanding scientometrics, their practical use in literature remains somewhat limited. In this study, we establish a joint coauthorship and citation network using preferential attachment. As papers get published, we update the coauthorship network based on each paper's author list, representing the collaborative team behind it. This team is formed considering the number of collaborations each author has, and we introduce new authors at a fixed probability, expanding the coauthorship network. Simultaneously, as each paper cites a specific number of references, we add an equivalent number of citations to the citation network upon publication. The likelihood of a paper being cited depends on its existing citations, fitness value, and age. Then we calculate the journal impact factor and $h$-index, using them as examples of scientific impact indicators. After thorough validation, we conduct case studies to analyze the impact of different parameters on the journal impact factor and $h$-index. The findings reveal that increasing the reference number $N$ or reducing the paper's lifetime $\theta$ significantly boosts the journal impact factor and average $h$-index. On the other hand, enlarging the team size $m$ without introducing new authors or decreasing the probability of newcomers $p$ notably increases the average $h$-index. In conclusion, it is evident that various parameters influence scientific impact indicators, and their interpretation can be manipulated by authors. Thus, exploring the impact of these parameters and continually refining scientific impact indicators are essential. The modeling and simulation method serves as a powerful tool in this ongoing process, and the model can be easily extended to include other scientific impact indicators and scenarios.

## 1. Introduction and Literature Review

Bibliometrics or scientometrics, the application of statistical or quantitative method to analyze scientific literature and discover 'laws' governing various quantities or indicators measuring research impact, is a well-established field. Over the past fifty years, extensive research has delved into the publication, citation, and clustering behaviors of scientific papers, revealing numerous quantitative 'laws'. Despite this, the utilization of mathematical modeling and computer simulations in bibliometric research remains infrequent and isolated, despite their potential advantages over traditional approaches [1]. For example, by replicating the microscopic behavior and coordination of researchers, as well as the structure and evolution of literatures, these methods can elucidate the underlying mechanisms behind certain macroscopic phenomena. Additionally, computer modeling and simulation offer a controlled setup and environment, circumventing biases and errors in real databases, allowing for a detailed assessment of the system's various aspects [2]. Furthermore, modeling and simulations often serve predictive purposes, constructing 'artificial societies' and conducting 'thought experiments' to explore extreme scenarios that are challenging to study in real life. The outcomes can inform decision-making and even policy design. In contrast to empirical data-oriented studies, these methods generate simulated data, facilitating a comparison with empirical results. Ref. [3] provides a comprehensive review of the mathematical models and simulation methods used in science, highlighting its cross-disciplinary nature involving scientometrics, network sciences, system sciences, computational sociology, and complexity theory. According to Ref. [3], science can be defined as a social network of researchers generating and validating a knowledge network. Consequently, coauthorship and citation networks are the most relevant, and this review will delve into them in detail.

In both citation and coauthorship networks, the Matthew effect, also known as preferential attachment or cumulative advantage, stands as the fundamental mechanism, signifying that success tends to breed more success [4]. This citation network mechanism found formal expression in a model first presented by de Solla Price in 1976. This model assumes a continuous publication of new papers, with the probability of citing a specific paper being proportionate to its existing number of citations [5]. Barabási later formalized these mechanisms, referring to them as 'growth' and 'preferential attachment', respectively [6]. However, Price's model, while capable of replicating the fat-tailed distribution of citations, assumes uniform paper quality and an inability for new papers to surpass older ones in citations, which is unrealistic. Consequently, the concept of 'fitness' is introduced to quantify each paper's inherent ability to accrue citations, leading to the development of fitness models, such as the Bianconi-Barabási model [7]. In the fitness model, papers published later with high fitness can outperform established citation leaders. Finally, as new ideas are integrated into the subsequent works, the novelty of each paper eventually diminished over time, a phenomenon known as the obsolescence or aging of scientific literature, studied since 1943 [8]. The aging term is often modeled with a negative exponential [9, 10] or log-normal shape [11]. Combining these four mechanisms (growth, preferential attachment, fitness, and aging) yields the minimal citation model that captures a paper's time evolution in citations [12]. Medo and Cimini [2] introduced a normalization term to counterbalance the undue advantage of early papers during the initial period. Concerning citation distributions across disciplines and years, it has been discovered that all these distributions can be rescaled on a universal curve when considering a relative indicator $c_f$, enabling the comparison of papers and authors from different fields [13].

The coauthorship network is often regarded as a social network, a topic extensively explored in the realm of social sciences. However, the application of statistics, mathematical modeling, and simulations to large-scale coauthorship networks for uncovering universalities and mechanisms has seen recent progress and has garnered less attention compared to citation networks [14]. In contrast to citation networks, where nodes represent papers and links are directed from new papers to existing ones, coauthorship networks feature authors as nodes and undirected links, allowing for the formation of new connections between existing nodes [15]. Newman [14, 16, 17] has studied the structures and statistical properties of coauthorship networks using empirical data from various bibliographic databases. However, Newman's focus is primarily on static networks. Tomassini [18] delve into the formation and temporal evolution of coauthorship networks by analyzing time-resolved empirical data. Additionally, Barabási [15] has proposed a mathematical model capturing the network's time evolution and used its results to elucidate empirical measurements. In Barabási's model, the coauthorship network continually expands through the addition of new authors and the incorporation of new internal links, representing papers co-authored by authors already in the database [15]. While Barabási's preferential attachment-based model effectively replicates and explains coauthorship networks, the utilization of parameters like average internal links $a$ and incoming links $b$ in unit time makes the connections between papers and authors less transparent. To address this issue, the mechanism of paper team assembly deserves scrutiny, as the paper team is the foundational element of the coauthorship network. A paper team consists of authors collaborating on a paper, forming a complete graph within the coauthorship network. Research indicates that paper team size gradually increases across all disciplines [19]. Guimera et al. [20] have explored the team assembly mechanisms determining the structure of collaboration networks, proposing a model for the self-assembly of creative teams based on three parameters: team size $m$, the fraction of newcomers in new productions $p$ and the tendency of incumbents to repeat previous collaborations $q$. Analyzing the team assembly mechanism could replace the parameters $a$ and $b$ in Barabási's model with more explicit ones $m$, $p$ and $q$.

Despite the potential insights gained from modeling the coevolution of citation and coauthorship networks, there has been a limited amount of research in this area [21]. One such model is the TARL (topics, aging, and recursive linking) model proposed by Börner [22] which posits that authors read and cite the references of randomly selected papers. Since highly cited papers are more likely to appear in the reference list, the TARL model successfully incorporates the Matthew effect into citations, leading to the reproduction of a fat-tailed distribution. However, a limitation of the TARL model is its assumption that each author produces a fixed number of papers per year, preventing it from accurately replicating the fat-tailed distribution of coauthors and the number of papers published per author. Another model, proposed by Xie et al. [21], employs a graphical model introducing new concepts like concentric circles, leaders, and influential zones. Xie's model considers the Matthew effect by acknowledging that older leaders, with larger influential zones, easily attract collaborators and publish more papers, consequently receiving more citations. While Xie's model successfully reproduces the fat-tailed distribution of citations and coauthors, it introduces an abundance of new concepts, assumptions, and parameters, making his graphical model less common in the literature.

Modeling and simulation methods are particularly well-suited for evaluating various scientific impact indicators, and there have already been studies in this area [2]. Among the scientific impact indicators, the journal impact factor [23] and the $h$-index [24] stand out as the most well-known and widely used. Garfield [23] first proposed the journal impact factor in 1972, defining it essentially as the average citations per published item. Zhou et al. [25] developed a citation model to investigate the impact of various publication factors, such as the average review cycle, the average number of references, and the yearly distribution of references, on the journal impact factor. Their findings suggest that journals with shorter review cycles, higher reference numbers, and more recent reference distributions tend to have higher impact factors [25]. In a subsequent study, Zhou et al. [26] expanded their approach to include a submission model that simulated the journal targeting, peer review, and publication processes. This extended model allows for the analysis of temporal dynamics and the distribution of impact factors across multiple journals.

The $h$-index, introduced by Hirsch [24] in 2005, serves as a composite metric reflecting both an author's productivity and impact, commonly utilized to gauge the accomplishments of a particular author. Guns and Rousseau [27] conducted an investigation through computer simulations of the publication and citation processes. Their findings indicate that, in most instances, the $h$-index exhibits linear growth over time, with occasional occurrences of an S-shaped pattern [27]. Ionescu and Chopard [28] contributed two agent-based models addressing performance measurements for individual scientists and groups of scientists. In their multi-scientist model, author productivity is assumed to follow Lotka's law, and distributions of the $h$-index are presented [28]. Medo and Cimini [2] conducted a comparative analysis of various scientific impact indicators, employing a citation model calibrated with empirical data. Assuming authors' productivity adheres to Lotka's law, they found that the $h$-index indeed captures the combined ability and productivity of researchers, although it may not consistently provide a fair comparison between researchers at different career stages [2].

## 2. Model Formulation and Validation

### 2.1 APS database

The coevolution model of coauthorship and citations relies on and is validated against the American Physical Society (APS) dataset, extensively employed in bibliometric studies [2, 9, 11, 29], and publicly available through Ref. [30]. The APS dataset comprises two subsets: citing article pairs and article metadata. Citing article pairs consist of pairs of APS papers, with one paper citing another, making them suitable for constructing citation networks. On the other hand, article metadata includes fundamental details like doi, authors, and publication dates for all APS papers, facilitating the construction of coauthorship networks. In this study, we exclusively consider citation pairs in which both the citing and cited papers fall within the article metadata subset. This choice ensures a consistent and precise match between total reference and citation numbers at all times.

The APS datasets cover materials from 1893 to the end of 2021, providing a continuous span of 129 years of empirical data. For the subsequent simulation, we have chosen a time length of $T = 13$ years, with each simulated year corresponding to approximately 10 years of empirical data. Although the APS datasets consist of 19 journals, this paper does not focus on comparing indicators across different journals. Instead, we treat the APS datasets as a unified virtual journal, where all references and citations occur within this virtual journal. Consequently, the simulation models only one journal with 12 issues per year.

## *2.2 growth of papers and authors*

The APS datasets comprise approximately 0.7 million papers and 0.5 million authors at the end of 2021. Figure 1(a) illustrates the annual growth of accumulated papers and authors. Utilizing the exponential growth model $P_t = \alpha \exp(\beta t)$ on the cumulative paper number from Figure 1(a), the estimated annual growth rate $\beta$ is determined to be 6.36 %. In this model, the initial year's 12 issues each contain $N_1 = 10$ papers, and with each subsequent year ($t$ increasing), $N_t$ increases by 1 paper. Consequently, by the end of the 13$^{th}$ year, each issue contains $N_{13} = 22$ papers. These issue arrangements correspond to an annual paper growth rate of 6.68 %, aligning closely with empirical results. Therefore, a total of $P = 2496$ papers will be modeled in the subsequent simulation.

Examining Figure 1(a), it becomes evident that the cumulative author number also follows an exponential increase over time. By plotting the cumulative author number against the cumulative paper number and performing linear fitting $y = kx$ for the data (Figure 1(b)), it is observed that, on average, with each new paper increment, approximately $k = 0.679$ new authors are added to the existing author list. Since each paper may involve multiple authors (e.g., $m$ authors), each author will be assessed independently whether he/she is a newcomer (with probability $p$) or an incumbent (with probability $1 - p$). As the number of newcomers within a single paper follows a binomial distribution, the expectation of newcomers for each paper is given by:

$$k = mp \tag{1}$$

where $m$ is the average team size ($m = 3.54$ for APS datasets). Consequently, the probability of selecting newcomers can be calculated as $p = k/m = 0.192$.

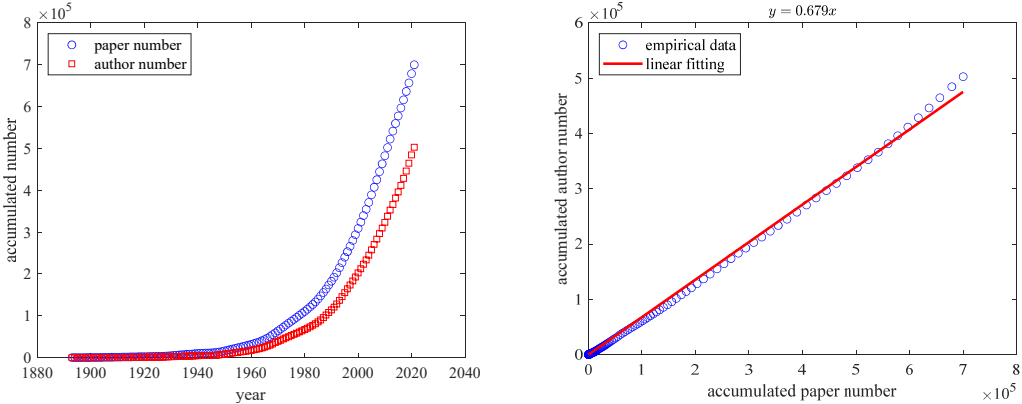

**Figure 1.** Evolution of cumulative papers and authors: (**a**) yearly progression; (**b**) author accumulation in relation to paper accumulation.

*2.3 paper team assembly*

A paper team refers to a collective of researchers who coauthor a paper, and their names collectively appear in the authors' field of a paper. Recent studies indicate that the average size of paper teams increases over time [19] and the distribution of paper team sizes exhibits a fat-tailed pattern [31]. This trend is similarly observed in the APS dataset, as indicated by the blue circles in Figure 2. As previously mentioned in Section 2.1, one simulation year corresponds to roughly 10 actual years of APS metadata. Consequently, the data on paper team sizes in the APS datasets are divided into 13 intervals based on their publication date. The team size distribution of the $i$-th interval ($i = 1,2, \ldots ,13$) is utilized to generate the distribution for the corresponding simulation year. The simulated results are represented by the red squares in Figure 2.

Observing Figure 2(a) it is evident that at any time, the average paper team size in the simulated results closely aligns with the empirical data, displaying identical distributions. The black dots represent the annual fluctuations in average paper team sizes in the APS empirical data. However, in Figure 2(b), a subtle distinction is noted in the distribution of the simulated results compared to the empirical one, indicating a higher occurrence of papers with smaller team sizes in the simulation. This discrepancy arises from the fact that the paper growth rate for each interval is $\beta_{10} = 10\beta = 63.6\ \%$, whereas the growth rate for each simulation year is only 6.68 %. Consequently, the distribution of empirical data is influenced more by the later intervals, resulting in higher proportions of papers with larger team sizes.

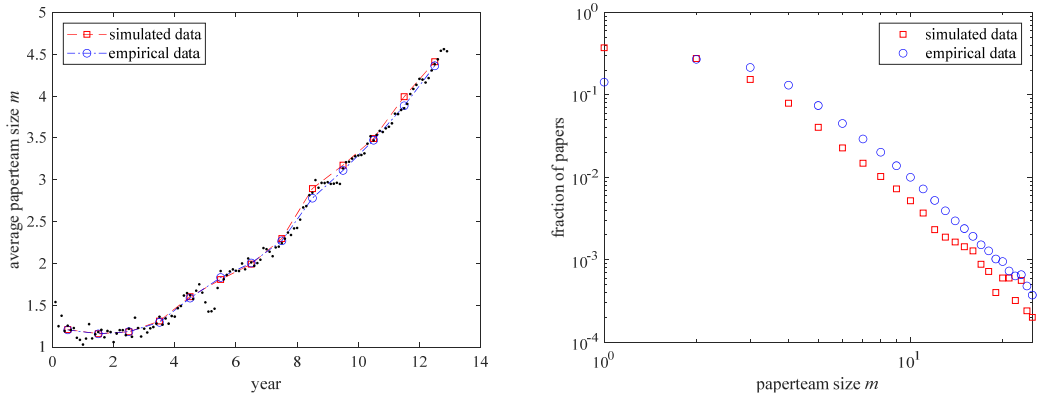

**Figure 2.** Model simulations vs APS empirical data: (**a**) annual average team size increase; (**b**) distribution of paper team sizes.

As mentioned in Section 2.2, the probability of selecting incumbents as authors is $1 - p$. If an incumbent is to be selected, the preferential attachment mechanism will be employed to determine which incumbent will be chosen. As demonstrated in Ref. [15] that the probability $\Pi(k)$ of a newcomer collaborating with an incumbent with connectivity $k$ can be expressed as $\Pi(k) \propto k^\nu$, where $\nu \leq 1$. Meanwhile, the probability $\Pi(k_1, k_2)$ of an incumbent with connectivity $k_1$ collaborating with another incumbent with connectivity $k_2$ can be factorizes into the product $k_1 k_2$, expressed as $\Pi(k_1, k_2) \propto k_1 k_2$. Therefore, in this simulation, the probability $\Pi(k)$ to select an incumbent with connectivity $k$ is set as:

$$\Pi(k) = (1 - p)\, k \Big/ \sum_{i \propto A_t} k_i \qquad (2)$$

where $p$ represents the probability of selecting newcomers, $k_i$ signifies the connectivity of each incumbent, and $A_t$ denotes the list of incumbents at time $t$. Given that incumbents often engage in repeated collaborations, characterized by the parameter $q$ in Ref. [20], the connectivity $k_i$ in this simulation refers to the accumulated number of collaborations rather than the number of collaborators an author has. For authors with no collaborations, an initial connectivity $k_0 = 1$ is assigned to ensure each author has a finite initial probability of being selected for the first time. Utilizing an adjacency matrix to record collaboration numbers for each pair of authors, the coauthorship network can be established, as further discussed in Section 2.5. In the event a new author is selected with probability of $p$, he/she will be added to the incumbent list $A_t$, and the adjacent matrix will be updated accordingly.

### *2.4 author ability and paper quality*

Recent research suggests that each scientist may possess a hidden intrinsic parameter, denoted as $Q$, which characterizes their ability to transform a random idea into works with varying impacts [29]. An author with a high $Q$-factor consistently experiences a successful career, regardless of the novelty of the projects or ideas they engage with [12]. The $Q$-factor has been demonstrated to be relatively independent of author productivity [29]. Consequently, when a new author publishes their first paper, a random $Q$-factor is assigned. In this simulation, a log-normal distribution with $\mu = 0.93$ and $\sigma = 0.46$ is assumed for the $Q$-factor, aligning with the data in Ref. [29], which is also based on the APS datasets. The distribution of authors' abilities ($Q$-factor) is depicted in Figure 3(a). As author ability is a continuous parameter, it is divided into 40 bins, and the binned results are illustrated as red squares in Figure 3(a).

Once a paper team $a_i$ is assembled, and the ability for each member $Q_j$ ($j \in a_i$) determined, the quality of the paper can be computed as $\eta_i = \delta \left( \max_{j \in a_i} Q_j \right)$, where $\delta$ represents a multiplicative noise term uniformly distributed in $[1 - \delta^*, 1 + \delta^*]$, introducing additional randomness to the paper creation process [2]. The distribution of the papers' quality is visualized in Figure 3(b), where a log-normal fitting is applied to the results, represented by the blue line in Figure 3(b).

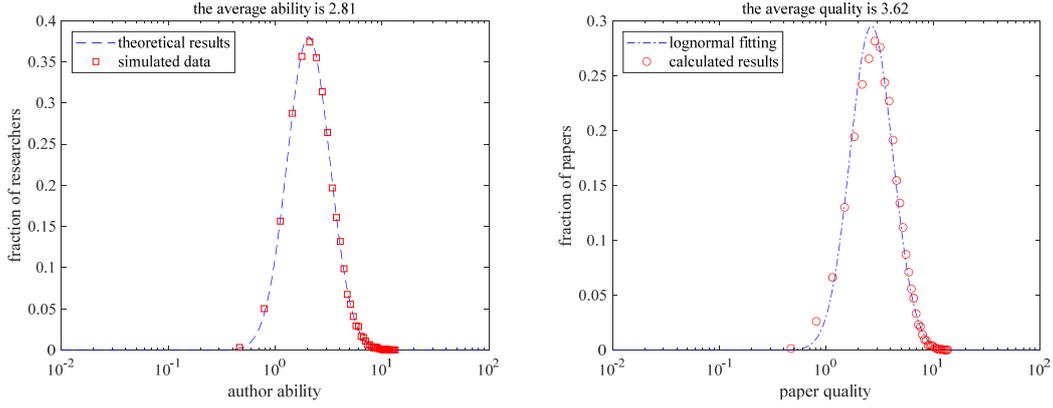

**Figure 3.** Distributions of author ability and paper quality: (**a**) author ability distribution; (**b**) paper quality distribution.

## 2.5 coauthorship network

Upon assembling a paper team, all its members essentially form a complete graph, prompting the update of the adjacent matrix of collaborations by incrementing each corresponding element by one. In this matrix $A$, each element $A_{i,j}$ represents the number of collaborations between author $i$ and author $j$. Simultaneously, the coauthorship network, essentially the collaborators' network, can be constructed by replacing the elements $A_{i,j}$ of the collaborations' network with 0 (for zero elements) or 1 (for non-zero elements). Additionally, the incumbents' list $A_t$ in Section 2.3 not only records the name or ID of an incumbent but also tracks the authored paper number (or productivity) of each author. Papers are incrementally added at each time step, and both the incumbents' list and coauthorship network evolve accordingly.

Once all $P = 2496$ papers are incorporated, the final distributions of productivity and collaborators are depicted in Figure 4. The productivity distribution essentially mirrors Lotka's law, as evident in Figure 4(a), where the simulated results closely align with empirical data. The distribution of collaborators in the simulated results also exhibits a strong match with empirical data, illustrated in Figure 4(b), thereby validating the coauthorship network model. Both distributions clearly display fat tails, and further discussions about the network of collaborators can be explored in Refs. [14-18].

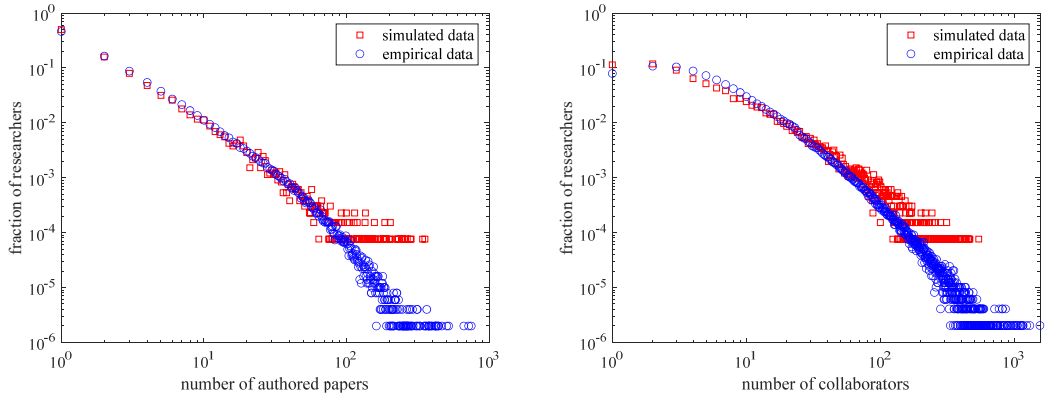

**Figure 4.** Model simulations vs. APS empirical data: (**a**) researcher productivity distribution; (**b**) collaborator number distribution.

*2.6 reference model*

As previously mentioned in Section 2.1, the total reference number precisely matches the total citation number at any given time in both the empirical dataset and the simulation. Consequently, since the average citation number gradually increases over time, the average reference number also experiences an upward trend, as depicted by the blue circles in Figure 5. Similar to the approach in Section 2.3, the reference number data for all papers are sorted based on their publication date and evenly divided into 13 intervals. The reference number distribution for the $i$-th interval is then utilized to generate the reference number distribution for the $i$-th simulation year ($i =$ 1, 2, ... , 13). The simulation results are represented by the red squares in Figure 5. As illustrated in Figure 5(a), the yearly average reference numbers in the simulated results closely align with empirical data, sharing identical distributions. However, Figure 5(b) indicates a subtle difference in the reference number distribution of all simulation data compared to the empirical one, with more papers exhibiting lower reference numbers. This discrepancy arises from the much higher paper growth rate for each interval (10 years) than that of each simulation year, leading the distribution of the empirical data to be influenced more by later intervals and consequently having more papers with higher reference numbers.

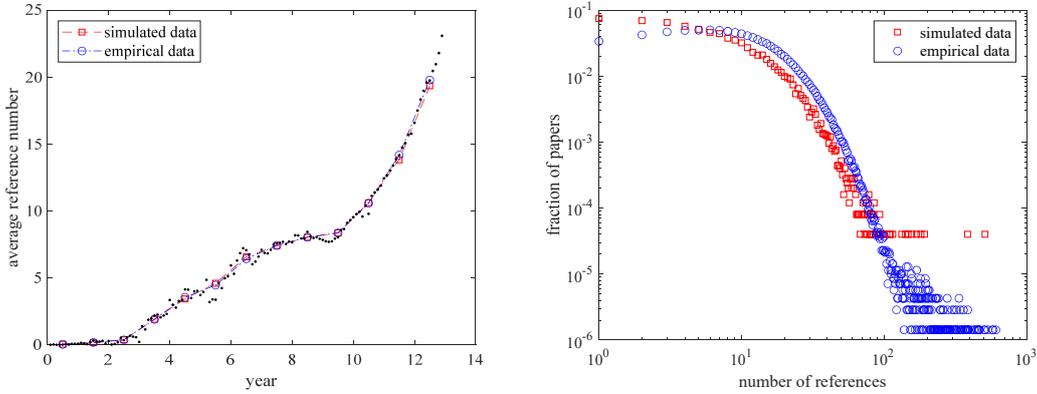

**Figure 5.** Model simulations vs. APS empirical data: (**a**) annual average reference numbers increase; (**b**) references number distribution.

*2.7 citation network*

Once the reference number for each paper is determined, the citation model can be established by determining which papers cite others. The citation network model utilized in this simulation is founded on the minimal citation model initially proposed by Wang et al. [11]. In this model, the probability that paper $i$ is cited at time $t$ after publication is determined by three independent factors: preferential attachment, fitness, and aging. The equation can be expressed as:

$$\Pi_i(t) = \eta_i c_i^t P_i(t) \qquad (3)$$

where $\eta_i$ represents the paper's fitness term, analogous to the paper's quality discussed earlier in Section 2.4, capturing the community's response to the work. $c_i^t$ is the preferential attachment term, indicating that the paper's probability of being cited is proportional to the total number of citations it has received previously. It's noteworthy that the preferential attachment term $c_i^t$ doesn't precisely equal the number of citations $n_{\text{cites}}(t)$. This is because we assign an initial attractiveness $c_0 = 1$ to a new paper with zero citations, ensuring each new paper has a finite initial probability of being cited for the first time [12]. Finally, the long-term decay in a paper's citation can be well approximated by a negative exponential aging term, expressed as $P_i(t) = \exp[-(t - \tau_i)/\theta]$, where $\tau_i$ is the publication date of the paper $i$, and $\theta$ is a parameter characterizing the lifetime of a paper [2]. In this paper, the value of $\theta$ is set to 48 months, consistent with the value employed by Refs. [2, 9], as their analyses are based on the same APS datasets.

The conclusive distribution of the citation network is depicted in Figure 6(a). Notably, it exhibits a fat-tailed pattern and aligns remarkably well with empirical data, thereby validating the citation network model.

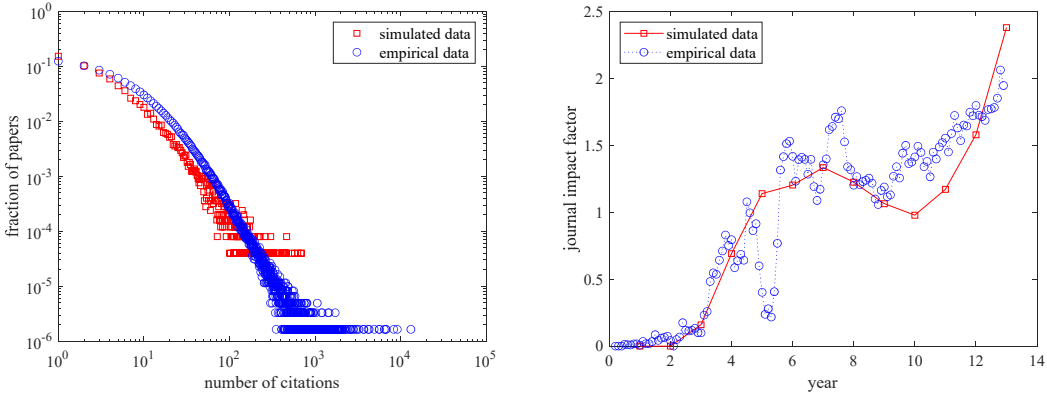

**Figure 6.** Model simulations vs. APS empirical data: (**a**) citation number distribution; (**b**) temporal variation of the journal impact factor of the APS dataset

### *2.8 journal impact factor*

The impact factor undergoes yearly fluctuations. When the counts of citations and papers are tallied from a given citation network, the yearly impact factor of a journal can be computed as follows:

$$IF(k) = \frac{n_{\text{cites}}(k, k-1) + n_{\text{cites}}(k, k-2)}{n_{\text{papers}}(k-1) + n_{\text{papers}}(k-2)} \quad (4)$$

where $IF(k)$ denotes the impact factor of the $k$th year; $n_{\text{papers}}(k-1)$ denotes the number of papers published in the $(k-1)$th year; $n_{\text{cites}}(k, k-1)$ denotes the number of those citations received during the $k$th year by the papers published in the $(k-1)$th year.

The fluctuation in journal impact factor is illustrated in Figure 6(b), where it can be observed that the simulated variations in journal impact factor closely align with the empirical results of the APS dataset, thereby validating the citation network model.

## 2.9 h-index

The $h$-index of an author is $h$ if $h$ of his papers have at least $h$ citations and each of the remaining papers have less than $h$ citations. To determine $h$, an author's publications are sorted based on their citations, arranged from the most cited to the least cited. This results in a sorted paper list, denoted as $\Pi = \{\alpha_1, \cdots, \alpha_i, \cdots, \alpha_n\}$ where $c_{\alpha_i} \geq c_{\alpha_{i+1}}$, $i \in [1, n-1]$. The $h$-index is then identified as the last position in which $c_{\alpha_i}$ is greater than or equal to the position $i$.

$$h = \max_i \left\{ \min_{\alpha_i \in \Pi} [c_{\alpha_i}, i] \right\} \tag{5}$$

The distributions and temporal variations of the $h$-index in both simulated and empirical results are illustrated in Figure 7. In Figure 7(a), it is evident that the $h$-index distributions for both simulated and empirical data exhibit fat-tailed characteristics and closely align with each other. These distributions, as depicted in Figure 7(a), also concur with the findings of Ref. [28], thereby validating the $h$-index outcomes from this simulation. Figure 7(b) presents the temporal dynamic growth of the top 3 researchers with the highest $h$-index. Notably, the general growth patterns in both simulated and empirical results are predominantly linear, consistent with the predictions in Ref. [27], adding credibility to the simulation results.

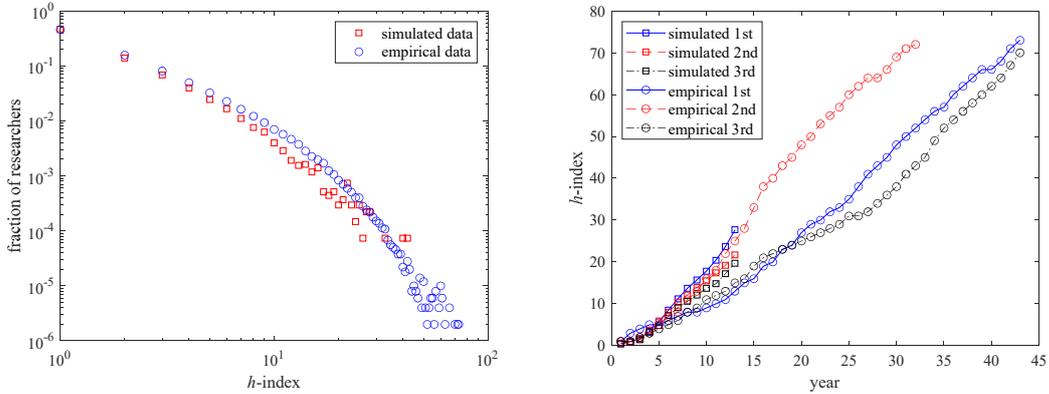

**Figure 7.** Model simulation versus APS empirical data: (**a**) $h$-index distribution in the final year; (**b**) temporal variation of the $h$-index for the top 3 researchers.

## 3. Results and Analysis

### 3.1 paper life time θ

The paper lifetime ($\theta$ in Equation (3)) is a parameter that characterizes the duration of a paper. For instance, if $\Delta t = t - \tau_i = \theta$, then $P_i(t) = 36.8\%$ in all cases. A larger $\theta$ implies that older papers will receive more citations. It is known that $\theta$ varies across different disciplines; for instance, mathematics tends to have a larger $\theta$ compared to biology. The effects of $\theta$ on the journal impact factor are depicted in Figure 8. It is evident that as $\theta$ increases, the journal impact factor decreases monotonically. This is because a larger $\theta$ results in more citations being attributed to older papers, particularly those published more than 2 years ago. Since the total number of citations remains constant, fewer citations are available for papers published within the last 2 years, which forms the numerator of Equation (4). Consequently, the journal impact factor decreases accordingly, as illustrated in Figure 8.

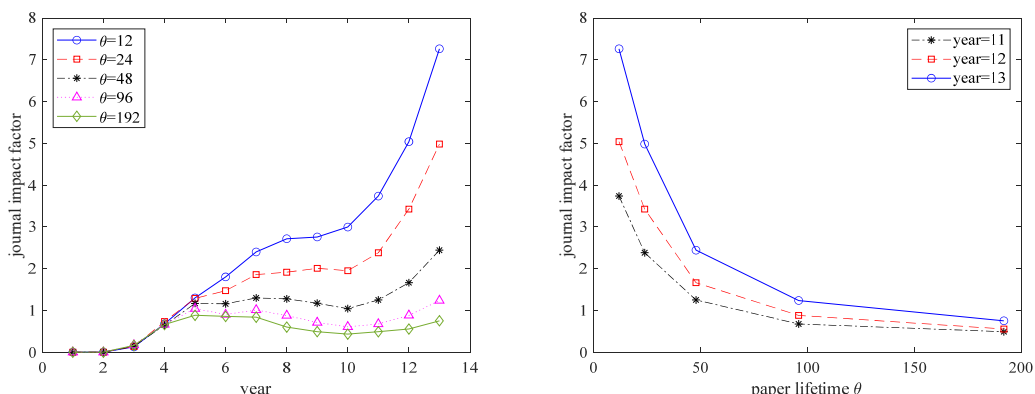

**Figure 8.** impact of paper life time $\theta$ on journal impact factor: (**a**) temporal variation of journal impact factor at different $\theta$; (**b**) the journal impact factor as functions of $\theta$ at different year.

The impact of paper life time $\theta$ on the distributions of $h$-index is illustrated in Figure 9(a). It is evident that a smaller $\theta$ corresponds to larger proportions of researchers with low or moderate $h$-index and smaller proportions of researchers with a large $h$-index. This is attributed to the fact that a small $\theta$ results in more citations being allocated to recently published papers, typically authored by newcomers with lower $h$-index. In contrast, a large $\theta$ leads to more citations directed at older papers, often authored by established incumbents, contributing to a stronger Matthew effect and resulting in a higher prevalence of researchers with large $h$-index, as observed in Figure 9(a). Researchers with lower or moderate $h$-index exhibit larger fractions, leading to a higher weighted average of distributions for smaller $\theta$, as depicted in Figure 9(b).

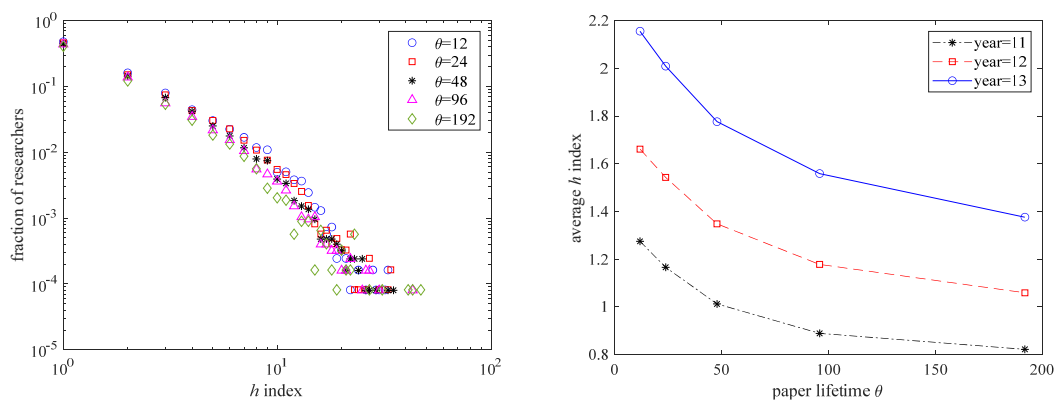

**Figure 9.** impact of the paper life time $\theta$ on the $h$-index: (**a**) distribution of $h$-index at different $\theta$; (**b**) average $h$-index as functions of $\theta$ at different year.

### *3.2 references number $N$*

As the average number of references $N$ equals the average number of citations, an increase in $N$ leads to higher $c_i^t$ in Equation (3) and consequently higher average citations. Given that the journal impact factor is directly influenced by the annual citations received by papers published in the recent 2 years, a higher $N$ results in a higher journal impact factor, as depicted in Figure 10.

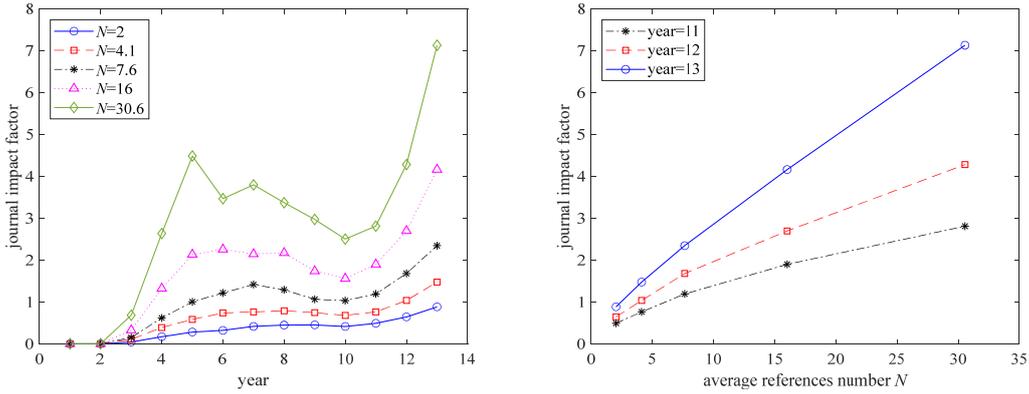

**Figure 10.** impact of reference number $N$ on journal impact factor: (**a**) temporal variation of journal impact factor at different $N$; (**b**) the journal impact factor as functions of $N$ at different year.

The $h$-index is influenced by both an author's productivity and the citations received by each paper. While increasing the average reference number $N$ has no direct impact on an author's productivity, it does contribute to increased citations for each published paper. Consequently, authors tend to have higher $h$-index values, as illustrated in Figure 11(a). The relationship between the average $h$-index of all authors and the reference number $N$ is depicted in Figure 11(b), where it is evident that the average $h$-index exhibits a monotonic increase with the reference number $N$.

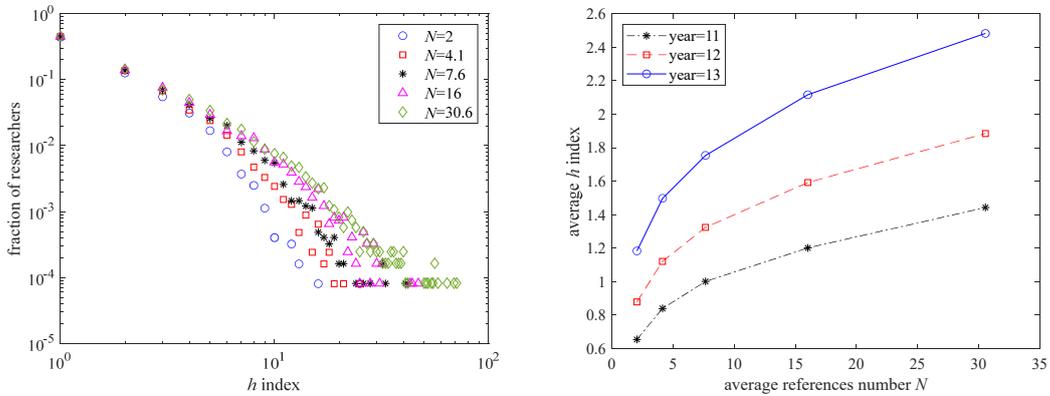

**Figure 11.** impact of the reference number $N$ on the $h$-index: (**a**) distribution of $h$-index at different $N$; (**b**) average $h$-index as functions of $N$ at different year.

### *3.3 team size $m$ at fixed $p$*

The impact of team size $m$ on the journal impact factor is minimal, as only the fitness term $\eta_i$ in Equation (3) is slightly influenced by $m$. Consequently, our analysis will primarily focus on how $m$ affects the distributions of the $h$-index, as depicted in Figure 12. With an increase in the average team size $m$, while keeping the probability of newcomers $p$ constant, more authors/researchers are generated with each published paper. Despite a larger $m$ resulting in each researcher being selected more frequently as a coauthor, the likelihood of getting selected each time decreases due to the higher number of researchers. Consequently, the average number of authored papers and the average $h$-index generally remain constant. Figure 12(a) indicates that with more researchers, the top researcher is more likely to achieve a higher $h$-index. However, since the total number of citations remains constant, fewer citations are available for the average researcher. As a result, the distributions of small team sizes tend to be higher than those of large team sizes in low to medium $h$-index region, as shown in Figure 12(a). As the average researcher occupies more fractions, the average $h$-index decreases with team size, as demonstrated in Figure 12(b).

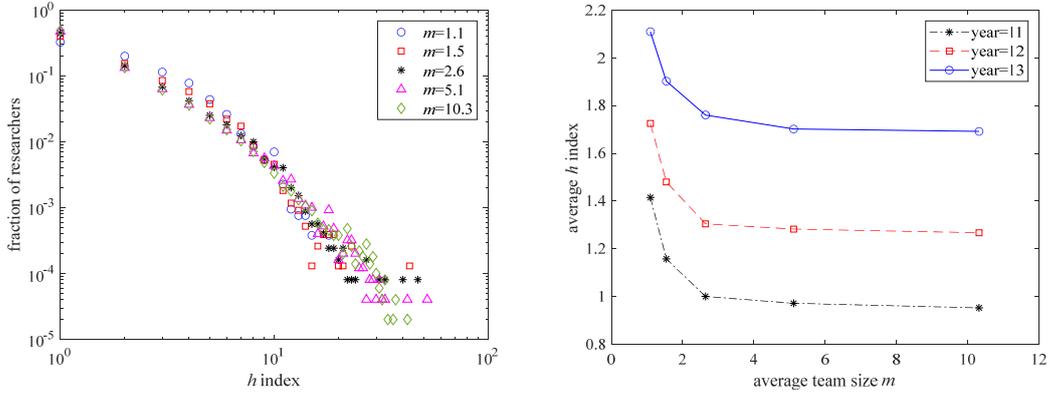

**Figure 12.** impact of the average team size $m$ on the $h$-index: **(a)** distribution of $h$-index at different $m$; **(b)** average $h$-index as functions of $m$ at different year.

### *3.4 probability of newcomers $p$*

Since the distributions of author's ability $Q$ are the same for newcomers and incumbents, the variation of $p$ has no impact on the paper's quality $\eta_i$ and thus does not affect the journal impact factor. When the probability of selecting newcomers $p$ increases while keeping the average team size $m$ constant, more newcomers are generated with each published paper and the probability of selecting incumbents as authors decreases. Therefore, as the probability of newcomers $p$ increases, the distributions of $h$-index will gradually become dominated by fresh researchers with low $h$-index, as shown in Figure 13(a). It can be noted that the distributions of small $p$ tend to be higher than those of large $p$. The average $h$-index will also decrease with the increasing $p$, as demonstrated in Figure 13(b).

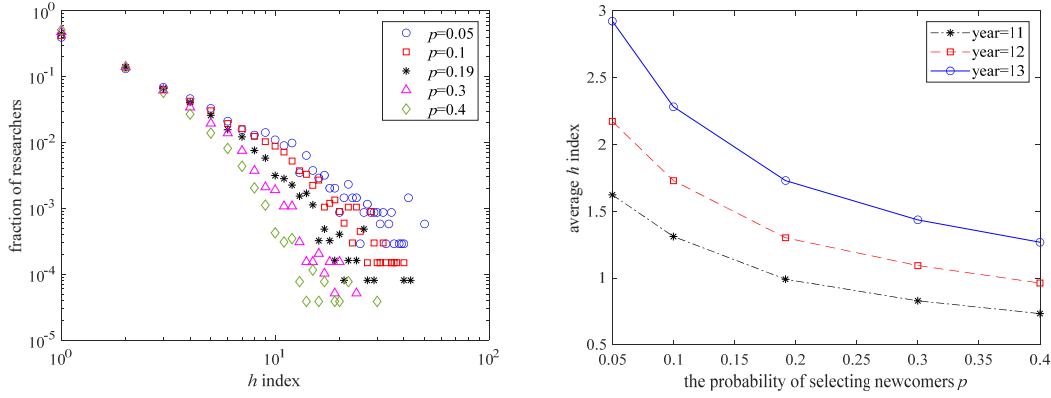

**Figure 13.** impact of the probability of newcomers $p$ on the $h$-index: (**a**) $h$-index at different $p$; (**b**) average $h$-index as functions of $p$ at different year.

### 3.5 team size $m$ at fixed $k$

In Section 3.3, we analyzed the impact of increasing team size $m$ while keeping the probability of selecting newcomers $p$ constant on the $h$-index. In this section, we will examine the effect of increasing team size $m$ while maintaining the number of new authors generated per each new paper $k$ constant. Since larger $m$ implies more frequent selections for each author, to uphold a constant $k$, the probability of selecting newcomers each time $p$ should be reduced accordingly. According to Equation (1), the probabilities of selecting newcomers are $p = [0.767, 0.384, 0.192, 0.096, 0.048]$ respectively when the team size is $m = [1.1, 1.6, 2.6, 5.2, 10.1]$. This case study simulates the scenario where incumbents intentionally enlarge their team size without the additional influx of newcomers. The impact of increasing $m$ while keeping $k$ constant on the distributions of $h$-index is shown in Figure 14(a). It can be noted that the numbers of authors with medium to high $h$-index increase significantly with the increasing team size $m$. This is because increasing the team size $m$ while keeping the new authors per paper $k$ constant can inflate the productivity of authors, especially those with more collaborations, and thus inflate their $h$-index. Consequently, the average $h$-index increases significantly with the increasing team size $m$, as shown in Figure 14(b).

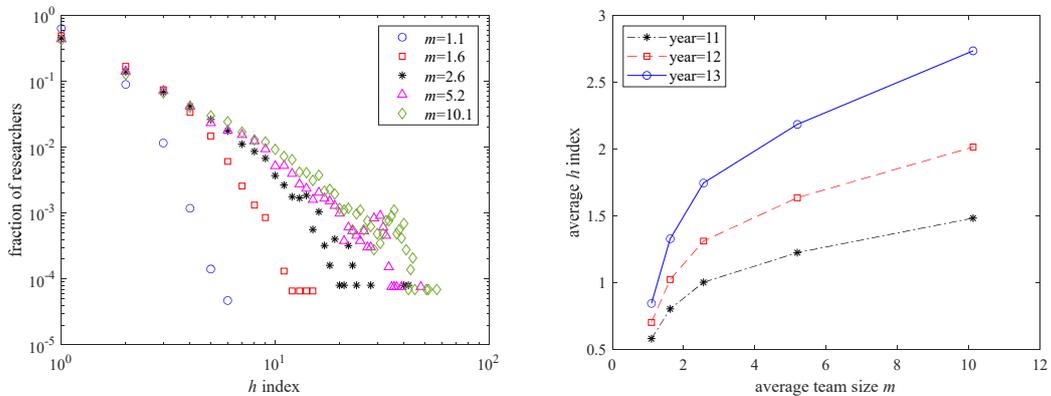

**Figure 14.** impact of the team size $m$ on the $h$-index: (**a**) distribution of $h$-index at different $m$; (**b**) average $h$-index as functions of $m$ at different year.

## 4. Conclusions

In this paper, mathematical model for the team assembly and citation process is established and the coevolution of coauthorship and citation network is simulated. Scientific impact indicators, such as the journal impact factor and $h$-index, are calculated and validated against the empirical data from the APS datasets. Parametric studies are conducted to analyze the impact of different parameters, such as the paper lifetime $\theta$, reference number $N$, team size $m$ and the probability of selecting newcomers $p$, on the journal impact factor and $h$-index. It can be concluded from this research that:

1. By using a few simple and reasonable assumptions, the mathematical models can effectively replicate most empirical data characteristics, including temporal dynamics and distributions of $h$-index, thus indicating that modelling and simulation methods are reliable tools for exploring how different parameters affect scientific impact indicators.

2. Increasing the reference number $N$ or decreasing the paper lifetime $\theta$ significantly boosted both the journal impact factor and average $h$-index. Additionally, enlarging team size $m$ without adding new authors or reducing the probability of selecting newcomers, notably increases the average $h$-index. This implies that scientific impact indicators may have inherent weaknesses or can be manipulated by authors, making them unreliable for assessing the true quality of a paper.

3. The presented mathematical models can be easily extended to include other scientific impact indicators and scenarios. This versatility positions modeling and simulation methods as powerful tools for studying the impact of various parameters on scientific impact indicators, aiding in the development of improved indicators. Furthermore, these methods can serve as robust tools for validating underlying mechanisms or predicting different scenarios based on joint coauthorship and citation networks.

## 5. Statements and Declarations:

**Funding and Conflicts of interests:** The research leading to these results received funding from the Library Society Guangdong under Grant Agreement No. GDTK23004. This article is one of the achievements of the 2023 Key Research Project of Guangdong Provincial Library titled "Joint Analysis and Data Governance of Papers and Patents in the Context of Smart Libraries" (Project Number: GDTK23004).## 6. References

[1]   GILBERT N. A simulation of the structure of academic science [J]. Sociological research online, 1997, 2(2): 91-105.

[2]   MEDO M, CIMINI G. Model-based evaluation of scientific impact indicators [J]. Physical Review E, 2016, 94(3): 032312.

[3]   SCHARNHORST A, BöRNER K, VAN DEN BESSELAAR P. Models of science dynamics:16